\documentclass[reprint,amsmath,amssymb,aps,prb,floatfix]{revtex4-2}
\usepackage{graphicx}% Include figure files
\usepackage{dcolumn}% Align table columns on decimal point
\usepackage{bm}% bold 
\usepackage{subfigure}
\begin{document}

%\preprint{APS/123-QED}

\title{Superconductivity in thin films of RuN}% Force line breaks with \\
%\thanks{A footnote to the article title}%

\author{A.S. Ilin}
 \altaffiliation[At present at ]{Department of Condensed Matter Physics, Weizmann Institute of Science, Israel.}%Lines break automatically or can be forced with \\
\author{A.O. Strugova}%
\author{I.A. Cohn}
\author{V.V. Pavlovskiy}
\affiliation{Kotelnikov Institute of Radioengineering and Electronics of RAS}

\affiliation{
 HSE University, Physics Department
}%
\author{A.V. Sadakov}
\affiliation{Lebedev Physical Institute of RAS}
\author{O.A. Sobolevskiy}
\affiliation{Lebedev Physical Institute of RAS}
\author{L.A. Morgun}
\affiliation{Lebedev Physical Institute of RAS}
\author{V.P. Matrovitskii}
\affiliation{Lebedev Physical Institute of RAS}
\author{G.V. Rybalchenko}
\affiliation{Lebedev Physical Institute of RAS}
\author{S.V. Zaitsev-Zotov}
 \email{serzz@cplire.ru}
\affiliation{Kotelnikov Institute of Radioengineering and Electronics of RAS}
\affiliation{
 HSE University, Physics Department
}%

%\date{\today}% It is always \today, today,
             %  but any date may be explicitly specified

\begin{abstract}
Superconductivity has been found in RuN films obtained by reactive magnetron sputtering. This is a novel member of the metal nitride superconductors family. The critical temperature of the superconducting transition varies depending on the substrate and ranges from 0.77~K to 1.29~K. The parameters of the crystal lattice of superconducting films have been determined: the lattice is distorted cubic with parameters $a=b=c=4.559$~\AA~for RuN/SiO$_2$, $a=b=c=4.536$~\AA~for RuN/Si, and $\alpha =\beta =\gamma =87.96^\circ$ for both films.  The upper critical magnetic field at zero temperature $H_{c2}(0)$ = 2.3–4.1~T and the coherence length $\xi=$~9–12~nm were found from experimental data using the WHH model, with $H_{c2}(0)$ is near the upper paramagnetic limit. An s-wave single band energy gap $\Delta(0) =  0.19$~meV was revealed by self-field critical current experiment at temperatures down to 10~mK.
\end{abstract}

%Use showkeys class option if keyword
                              %display desired
\maketitle

%\tableofcontents

\section{Introduction}

Transition metal nitrides have been attracting the interest of researchers for decades due to a wide variety of their electronic properties \cite{Wang2021_TMNreview}. Moreover, many of metal nitrides are superconductors with a relatively high critical transition temperature $T_c$, which turns out to be higher than the critical temperature of the initial transition metal element \cite{MN}. The superconducting properties of niobium and titanium nitrides are best studied, and to a lesser extent nitrides of most other transition metals: zirconium, vanadium, hafnium, molybdenum, tungsten, tantalum and rhenium. At the same time, due to the complexity of synthesis \cite{Gregoryanz2004}, the superconducting properties of nitrides of platinum group metals remain poorly understood. 

RuN films can be obtained by various methods: laser ablation of ruthenium in a nitrogen atmosphere \cite{RuN_PLD}, reactive magnetron sputtering \cite{Liao_2009,Wu_2011,Bouhtiyya_2013,Cataruzza_2014,Cataruzza_2016}. The material is characterised by a positive enthalpy of formation. As a result, it is not stable against heating and loses nitrogen when heated above 200$^\circ$C \cite{Wu_2011,Cataruzza_2016}. Consequently, no nitrogen is detected in films grown by reactive magnetron sputtering if the substrate temperature exceeds 100$^\circ$C \cite{Cataruzza_2016}. NaCl-like structure was reported for films deposited by pulsed laser deposition \cite{RuN_PLD} and ZnS-like structure for ones grown by magnetron sputtering \cite{Bouhtiyya_2013,Cataruzza_2014}.

To the best of our knowledge, there is no information on the superconductivity of RuN films in the literature. In this paper, we demonstrate the presence of superconductivity in RuN thin films and present the main characteristics obtained by studying the crystal structure using X-ray diffraction analysis, elemental composition using energy-dispersive spectroscopy, energy structure using density functional theory, and superconducting state parameters derived from measurements of critical temperature and current, depending on the magnetic field.

\section{Film Growth, Their Structure and Composition}
\subsection{Film growth}
RuN films were grown by reactive magnetron sputtering in a pure nitrogen atmosphere on substrates of various types: single-crystal silicon (Si), thermally oxidized single-crystal silicon (Si/SiO$_2$), and quartz glass (SiO$_2$). Film deposition was carried out in a VON ARDENNE LS 730S setup. The residual gas pressure in the chamber and the nitrogen pressure were $< 10^{-7}$ and $6\times 10^{-3}$~mbar, respectively, at a discharge current of 70-80~mA. The typical film thickness was around 100 nm. The films under study were deposited at room temperature in a single technological circle.

\subsection{Film composition}
\begin{figure}[ht]
\includegraphics[width=7cm]{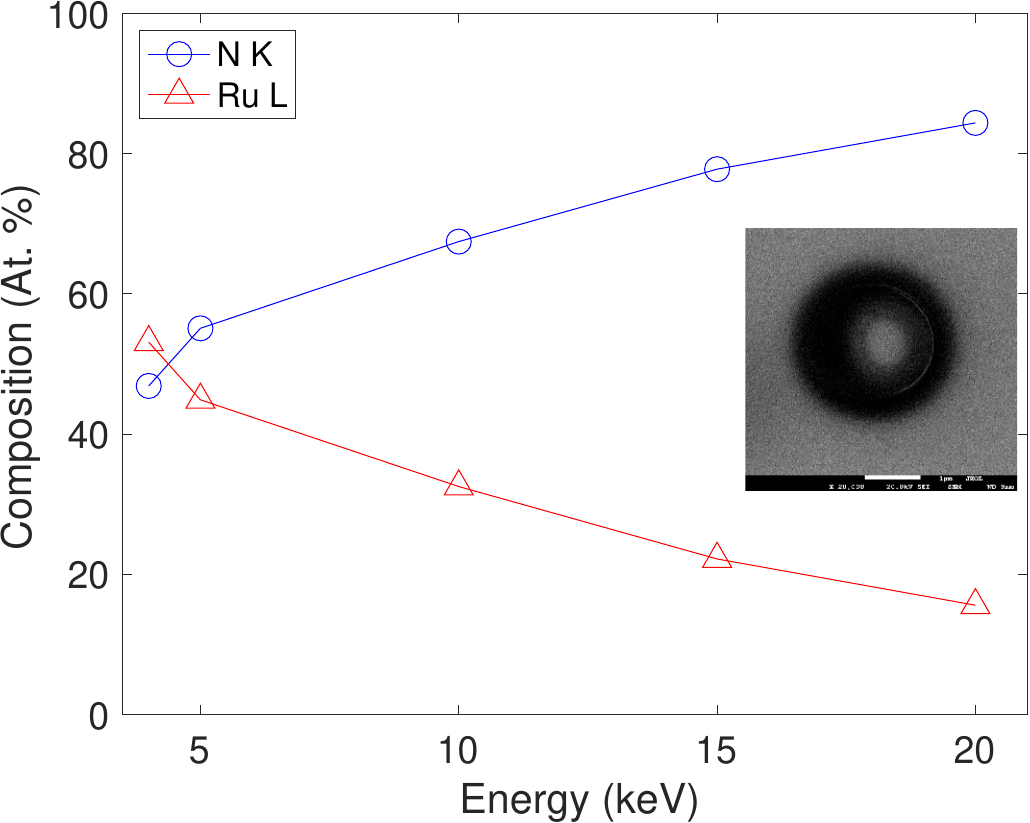}
\caption{\label{fig:comp} EDS results obtained at different electron energies in RuN/Si film. The inset shows the area under analysis damaged by an electron beam with an energy of 20 keV. Scale mark is 1 $\mu m$}
\end{figure}

Figure~\ref{fig:comp} shows the elemental composition of a RuN/Si film obtained by the energy-dispersive X-ray spectroscopy (EDS) at different electron energies. We see that EDS results depends on the e-beam energy.  This dependence corresponds to a shift of the effective analysis area deeper into the substrate with the e-beam energy increase (the substrate contribution at 20 keV is dominant) and temperature instability of the films around 200$^\circ$C observed earlier \cite{Wu_2011,Cataruzza_2016}. The damage produced by e-beam during collecting EDS data at large energies is clearly seen in SEM images of the film surface (see inset in Fig.~\ref{fig:comp}). 
The study of the elemental composition of the RuN films, the superconducting properties of which were investigated, carried out at an electron energy of 4~keV at which the substrate contribution does not exceed a few percent, gave 49.7\%, 51,1\% and 50.6\% of nitrogen and 50.3\%, 48.9\% and 49.4\% of ruthenium contents in the films grown on Si, oxidized Si and SiO$_2$ substrates respectively.

\subsection{Films structure study}
Structural studies were carried out on a Panalytical MRD diffractometer with a parabolic X-ray mirror as the primary monochromator and a third parallel analyzer. The samples were attached to a crystal holder made of single-crystal silicon to reduce the background.  

\begin{figure}[htbp]
\includegraphics[width=8cm]{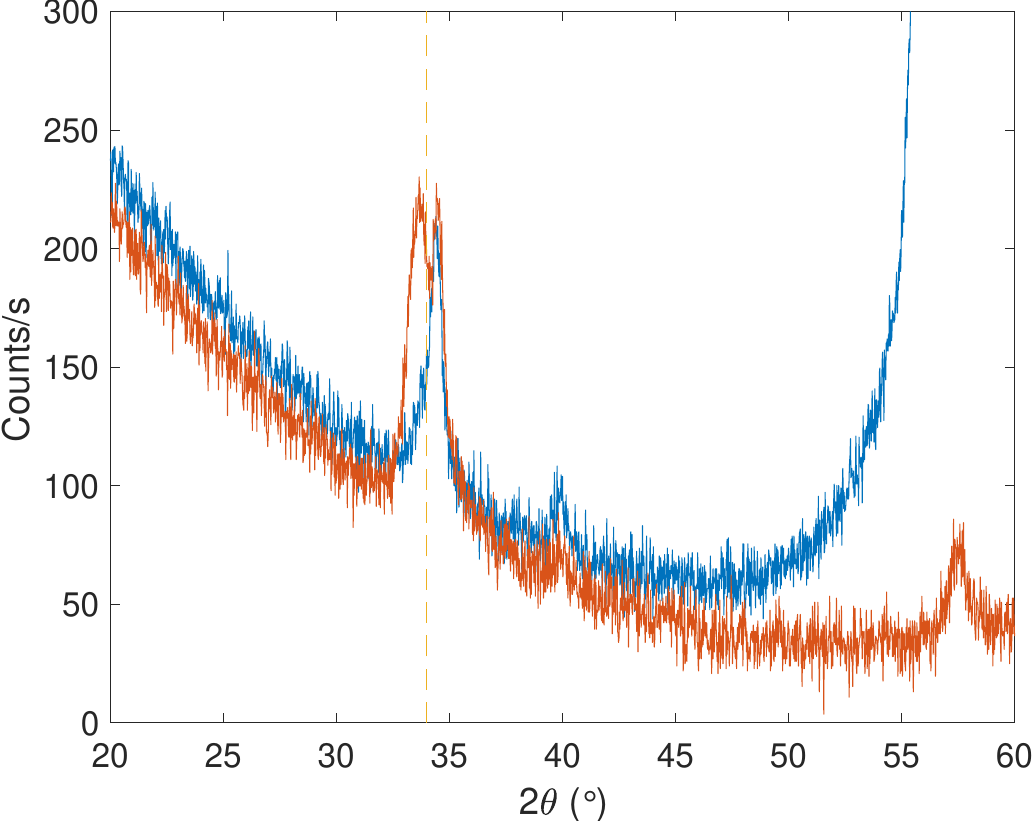}
\caption{\label{fig:Xray1} Square scale X-ray diffraction patterns of a RuN sample on silicon obtained with a joint $(2\theta$–$\omega)$-scan (blue) and with a fixed sample at an angle of $\omega =3^\circ$ for $2\theta$ scans (brick red).}
\end{figure}

Figure~\ref{fig:Xray1} shows square scale X-ray diffraction patterns of a RuN film on a silicon substrate obtained with a joint $(2\theta$–$\omega)$-scan (blue) and with a fixed sample at an angle of $\omega =3^\circ$ for $2\theta$ scans (brick red). In the first case (blue), the reflecting planes are parallel to the substrate surface; in the second case, they are inclined to the surface (brick red). In the range of angles $2\theta\approx 34^\circ$ on the red curve, one of the two peaks disappears whereas the intensity of the peak at an angle of $2\theta \approx 40^\circ$ on the red curve increases, which indicates the existence of a preferential grain orientation parallel to the substrate surface. Unfortunately, an asymmetric silicon peak with a sharp increase in intensity partially overlaps with the red curve.

\begin{figure}[ht]
\includegraphics[width=8cm]{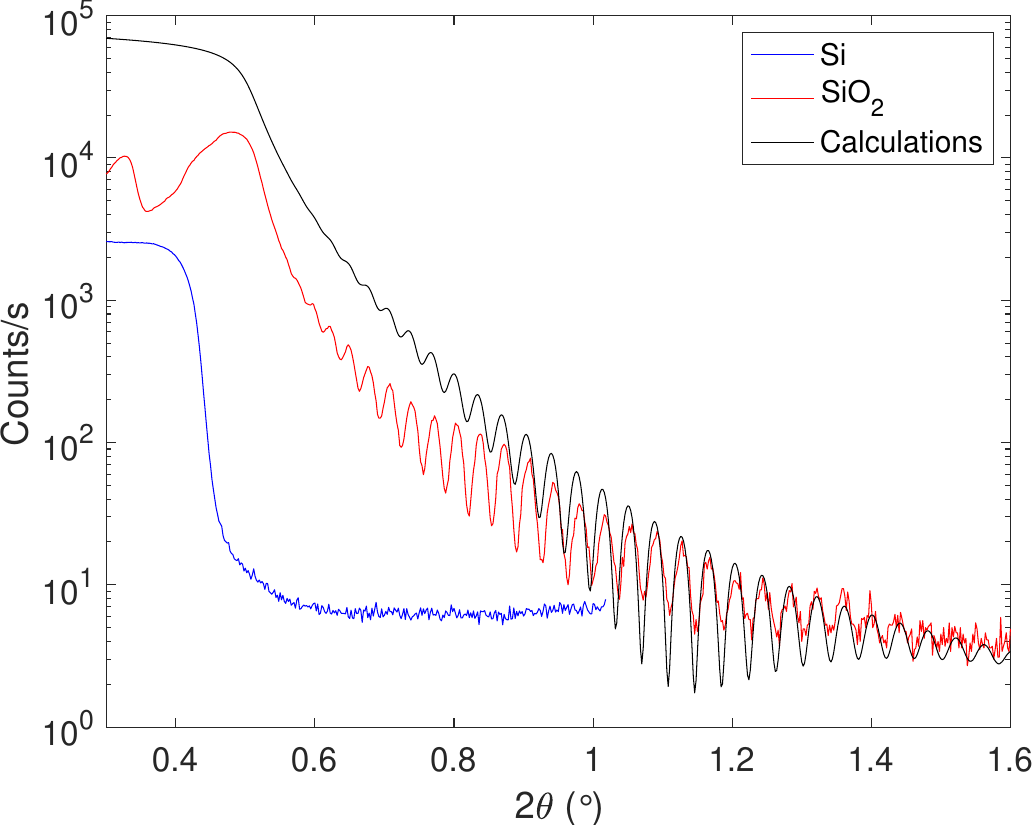}
\caption{\label{fig:Xray23} Reflectometric curves of RuN films on silicon (blue), SiO$_2$ (red) substrates and calculated curve (black) for a film with a thickness of 105 nm and a density of 7.5 g/cm$^3$ (black).}
\end{figure}

Figure~\ref{fig:Xray23} shows reflectometric curves of RuN films on silicon (blue) and SiO$_2$ (red) substrates. Ruthenium has a very high density (12.41 g/cm$^3$) compared to silicon (2.328 g/cm$^3$) or SiO$_2$ (2.64 g/cm$^3$). The greater the density, the greater the value of the angle of total external reflection, the greater the difference between the densities of the layer and the substrate, and the greater the intensity of the satellites. In RuN/Si, the density is noticeably lower than in RuN/SiO$_2$ where the experimental reflectometric curve agrees with the calculated one (black). In addition, satellites are completely absent on the blue curve, and the decrease in intensity with increasing diffraction angle corresponds to a roughness of more than 10 nm. Reflectometry does not distinguish between the roughness of the upper surface and the diffusion smearing of the interface between the substrate and the layer. However, a decrease in density unambiguously testifies in favor of strong diffusion mixing of the composition of the epitaxial layer with the substrate.

\begin{figure}[ht]
\includegraphics[width=8cm]{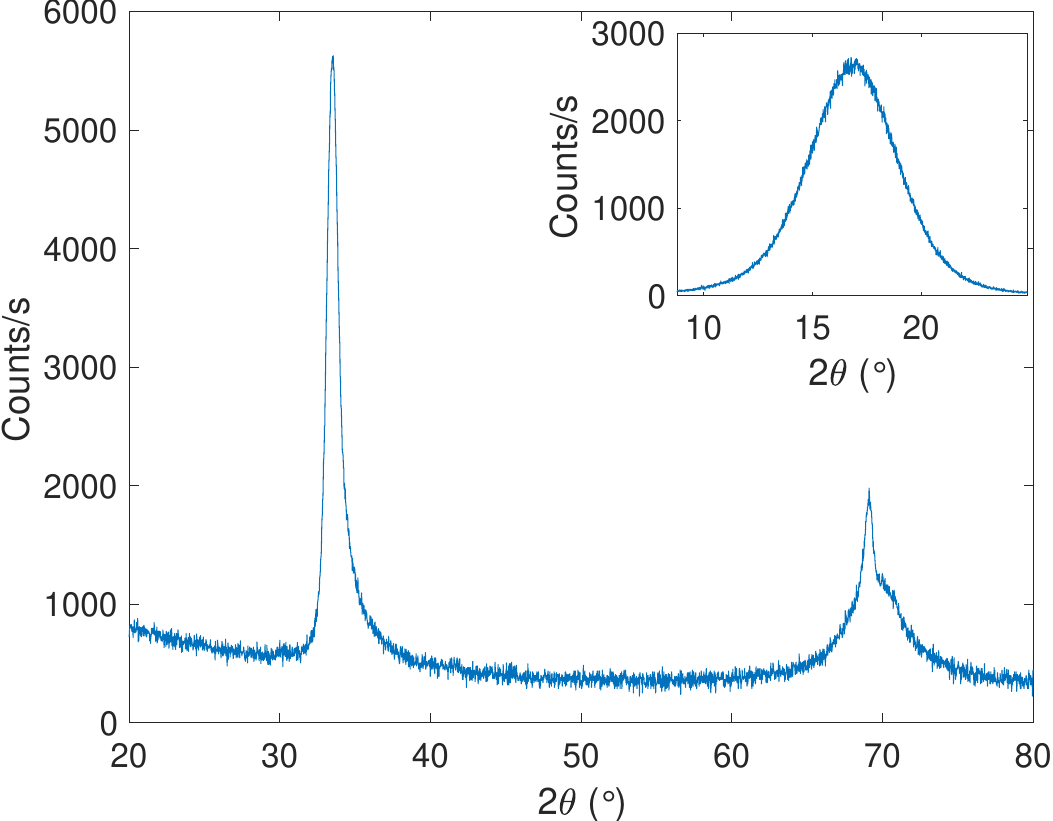}
\caption{\label{fig:Xray45} Diffractogram of the RuN/SiO$_2$ film, consisting of two orders of the same reflection. The inset shows rocking curve of the RuN/SiO$_2$ film with half-width $\Delta\omega=4.7^\circ$ for the first reflection in the diffraction pattern.}
\end{figure}

Figure~\ref{fig:Xray45} shows a diffractogram of the RuN/SiO$_2$ film. It consists of two orders of the same reflection and clearly indicates a tendency for the material to orient itself with its basal plane along the surface of the substrate. 

The rocking curve of the RuN layer on SiO$_2$ for the first reflection in the diffraction pattern is shown in the inset in Fig.~\ref{fig:Xray45}. The maximum intensity coincides with the diffraction angle. This means that the epitaxial layer grows on an amorphous substrate with a basal plane nearly parallel to the substrate plane. In other words, it is an axial texture in which the normals to the basal plane of individual crystallites are approximately parallel, and the remaining normals to the asymmetric reflecting planes of the crystallites are evenly distributed circularly.

First of all, it is desirable to understand a syngony RuN structure obtained and to calculate the unit cell parameters.
Ruthenium is better known in the hexagonal lattice $a_{Ru}=2.724$~\AA, $c_{Ru}=4.332$~\AA , but there is also a cubic variant $a_{Ru Cub}=3.83$~\AA .

Let's start with the more common hexagonal structure, especially since there is a tendency for the layer to crystallize with the basal plane parallel to the substrate. If nitrogen does not change the hexagonal structure, then the first reflection perpendicular to the basal plane has the index (0002) allowed for the hexagonal structure and the value of the lattice parameter along the $c$ axis is $c=5.350$~\AA . Then, the strong reflection with $d_3=2.612$~\AA \ located near the first reflection can be the reflection $(10\overline{1}1)$ with $a_1=3.459$~\AA , but the slope angle for it according to the crystallography formulas is $\phi_{(0001)/(10\overline{1}1)}=60.76^\circ$ instead of the experimentally observed angle of $70.8^\circ$. For reflection $(11\overline{2}1)$ with $a_2=5.992$~\AA \ $(0001)/(11\overline{2}1)=60.75^\circ $. No other reflections of the hexagonal phase are suitable for explaining the observed experimental data.

If the layer lattice is cubic, then the first symmetric reflection has indices (111) and $a_{cub}=4.633$~\AA \ for $d_{(111)}=2.675$~\AA . In a cubic lattice, there is only one family of planes $(\overline{1}11)$, $(1\overline{1}1)$ and $(11\overline{1})$, forming an angle of $70.53^\circ$ with the (111) plane, with the same interplanar spacing, and hence the same diffraction angle. But the experimental value of the interplanar distance for asymmetric reflections is smaller since the first pair of strong reflections on the diffraction pattern is bifurcated ($d_1=2.675$~\AA , $d_3=2.612$~\AA ). A possible explanation is the following: the cubic lattice undergoes rhombohedral deformation (the cube lattice is elongated along one spatial diagonal [111], as a result of which the length of the lattice changes slightly, and the angles of the unit cell become less than $90^\circ$). Let us find the magnitude of the deformation of the corners from the bifurcation of the first two reflections. For $d_1=2.675$~\AA \ and $d_3=2.612$~\AA \ in terms of interplanar distances and the angle between them, they correspond to the lattice $a=b=c=4.559$~\AA , $\alpha =\beta =\gamma=87.96^\circ$. The calculated angle of inclination of the three planes $(\overline{1}11)$, $(1\overline{1}1)$ and $(11\overline{1})$ to the (111) plane is $71.7^\circ$, which is close to the experimental value of $70.8^\circ$, taking into account the grain misorientation half-width of $4.7^\circ$. Let's check this conclusion on other reflections.

\begin{figure}[ht]
\includegraphics[width=8cm]{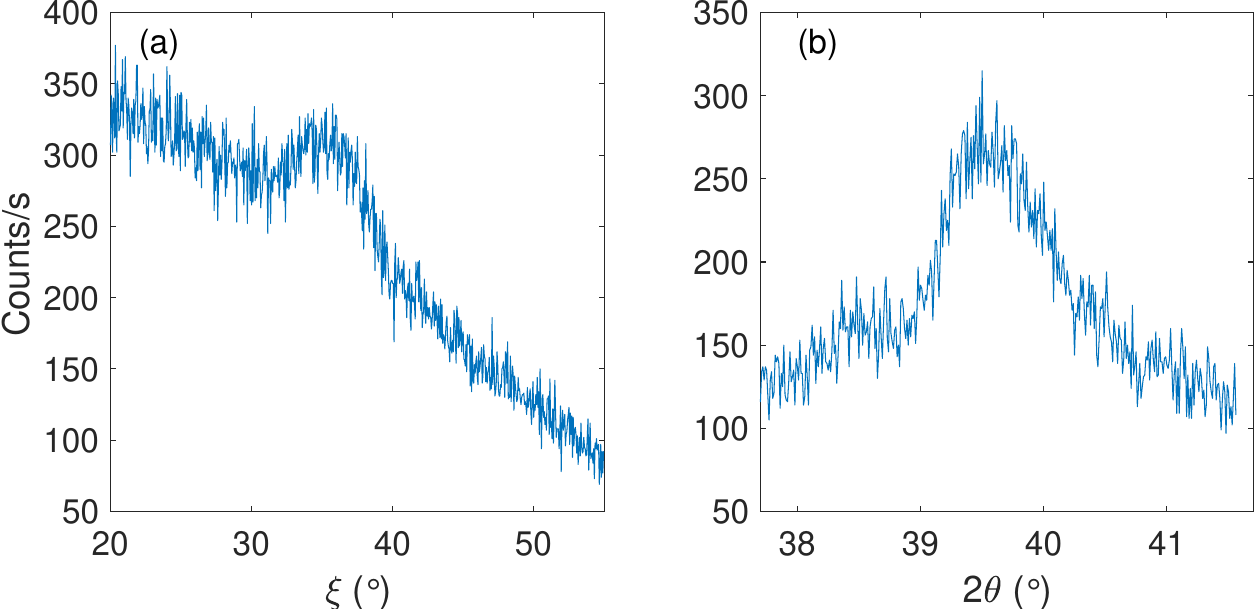}
\caption{\label{fig:Xray1213} a) Results of $\xi$-scanning when the sample plane deviates from the vertical with the angular position of the detector $2\theta =56.8^\circ$. The maximum intensity of the reflection is at $\xi=35.24^\circ$.
b) Results of ($2\theta$–$\theta)$–scanning on the asymmetric reflection of the RuN/SiO$_2$ film with the deviation of the sample plane by $35.24^\circ$ from the vertical. By the angular position $2\theta =56.764^\circ$ we find the value $d_5=1.6205$~\AA . For RuN/Si film, $d_5=1.6015$~\AA , which is noticeably smaller.}
\end{figure}

The reflection in Figs.~\ref{fig:Xray1213}(a),(b) with an angle of inclination to the basal plane of $35.24^\circ$ and an angular position of $2\theta =56.764^\circ$, $d_5=1.621$~\AA \ can be identified as the reflection (022) with calculated $d_{(022)}=1.624$~\AA \ and an inclination angle of $38.7^\circ$, while for an undeformed cubic lattice the inclination angle is $35.26^\circ$. Little discrepancies can be associated with the layer defects, which concentration is noticeably greater in the layer on silicon, due to which the maxima of the far reflection on the two substrates noticeably diverge ($2\theta_{RuN/Si}=57.50^\circ$, $2\theta_{RuN/SiO2}=56.764^\circ$).

Thus, all RuN reflections observed on the diffractograms belong to a cubic lattice deformed along one spatial diagonal, with parameters $a=b=c=4.559$~\AA , $\alpha =\beta =\gamma=87.96^\circ$ for RuN/SiO$_2$ film. Similarly, for RuN/Si film, the lattice parameters are  $a=b=c=4.536$~\AA , $\alpha =\beta =\gamma=87.96^\circ$.

\section{Numerical simulation of electronic band structure}
\begin{figure}[ht]
\includegraphics[width=8cm]{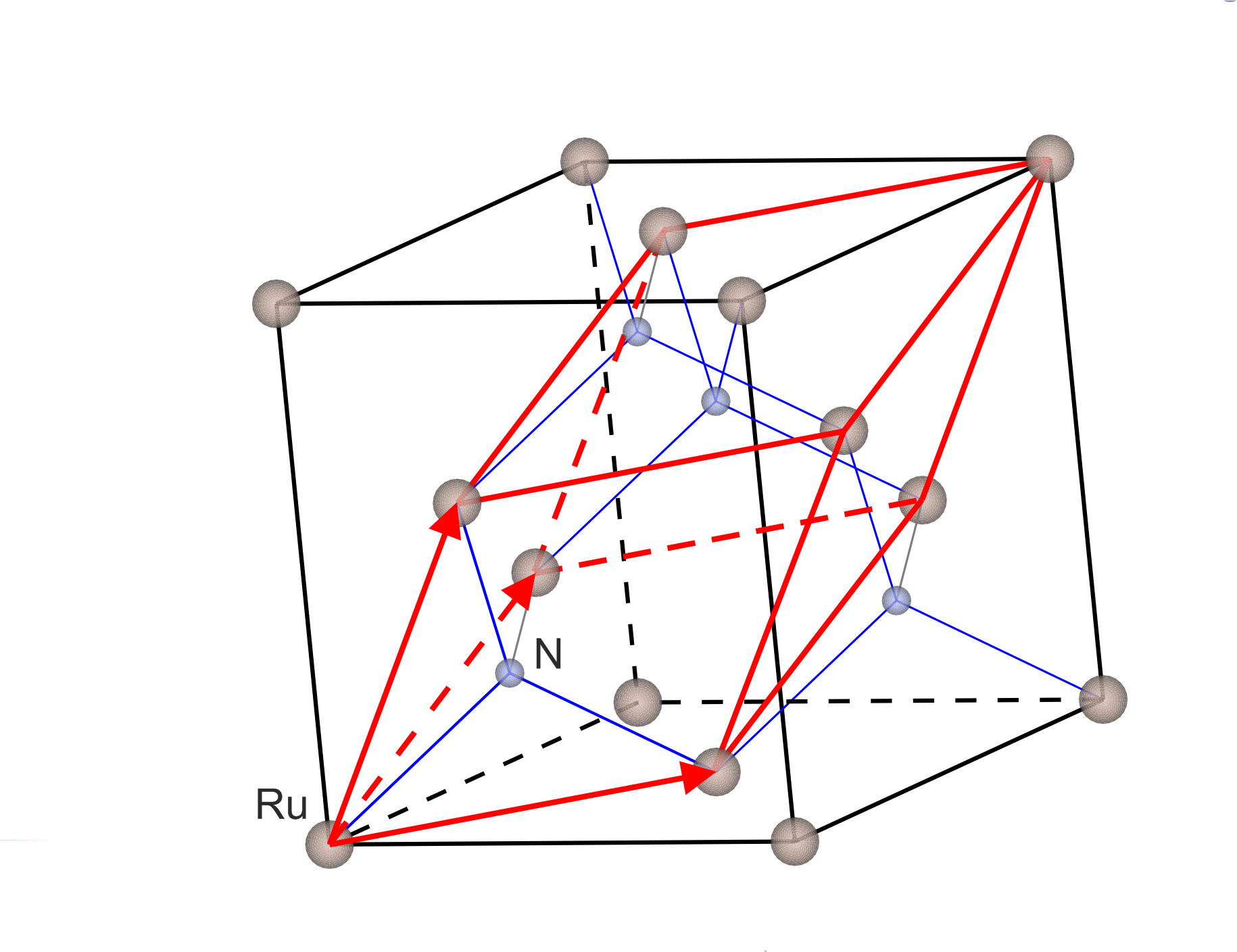}
\caption{\label{fig:UnitCell} Conventional (black) and primitive (red) unit cells of RuN.}
\end{figure}

The electronic band structure and Fermi surface calculations were performed using the Abinit package \cite{GONZE2020107042}. Conventional unit cell with the R3m:H orthorhombic symmetry \cite{Crystals2018} obtained above can be reduced to primitive unit cell with parameters $a = b = c = 3.2803$~\AA \ and $\alpha = \beta = \gamma = 57.70^\circ$ (Fig.~\ref{fig:UnitCell}). Structure optimization was performed with full unit cell optimization. 
Resulting lattice vector lengths were reduced by 0.05 \AA ~(1.5\%) and the angles were increased by $2.3^\circ$ (4\%).
We used original lattice parameters obtained from the experiment for the electron band structure calculations.

 The calculations were performed in the framework of density functional theory (DFT) and based on exchange-correlation functional with the Perdew-Burke-Ernzerhof (PBE) parameterization of the generalized gradient approximation (GGA).  Spin-orbit coupling (SOC) was taken into account, and non-collinear full relativistic pseudopotentials were used because of the presence of rather heavy element Ru. An energy cutoff of 600 eV and $k$-point grid $31\times 31\times 31$ within the Monkhorst-Pack scheme were adopted for the Brillouin zone sampling.  
 
\begin{figure}
\includegraphics[width=8cm]{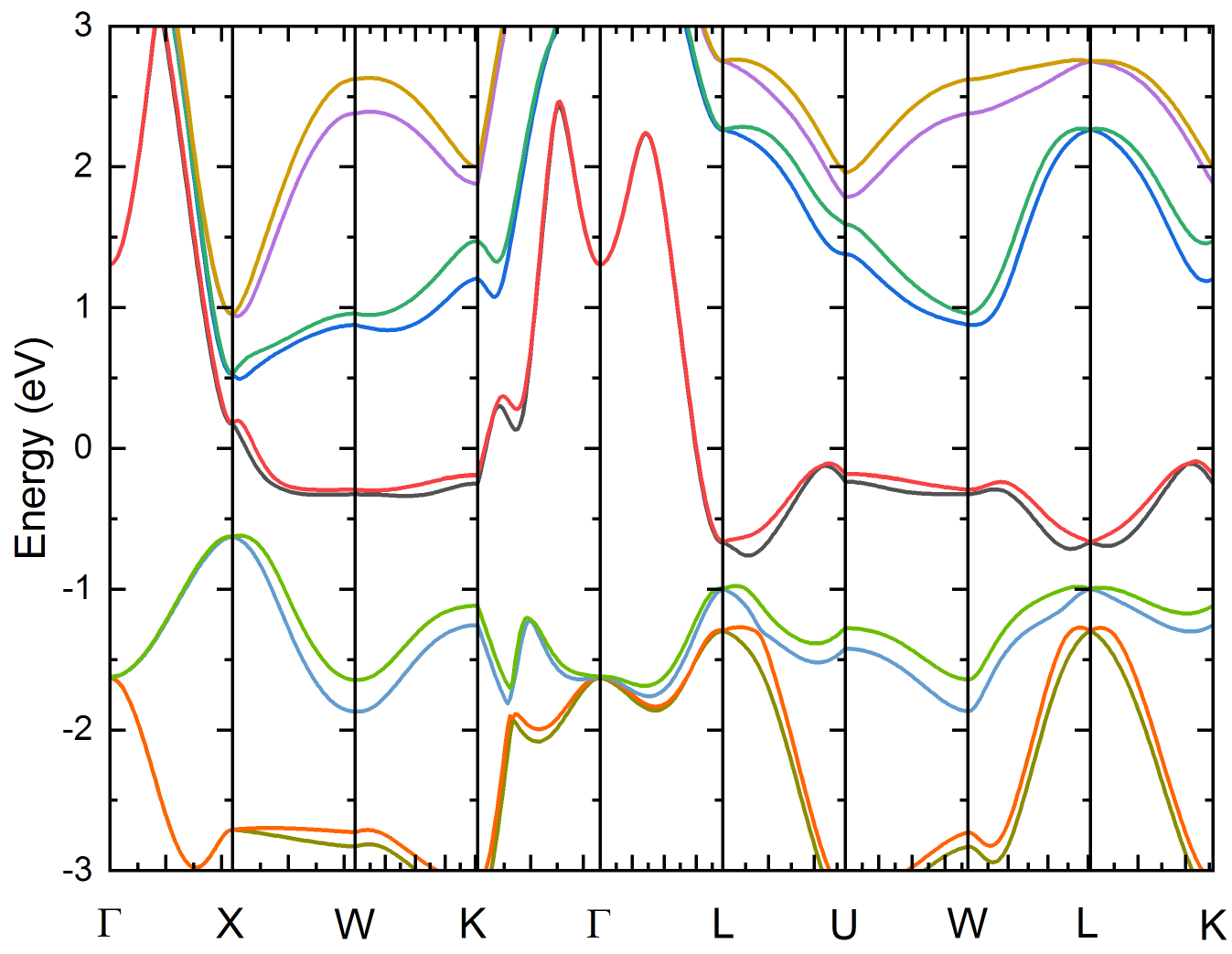}
\caption{ \label{fig:bands} RuN band structure.  }
\end{figure}

\begin{figure}
    \centering
    \includegraphics[width=5cm]{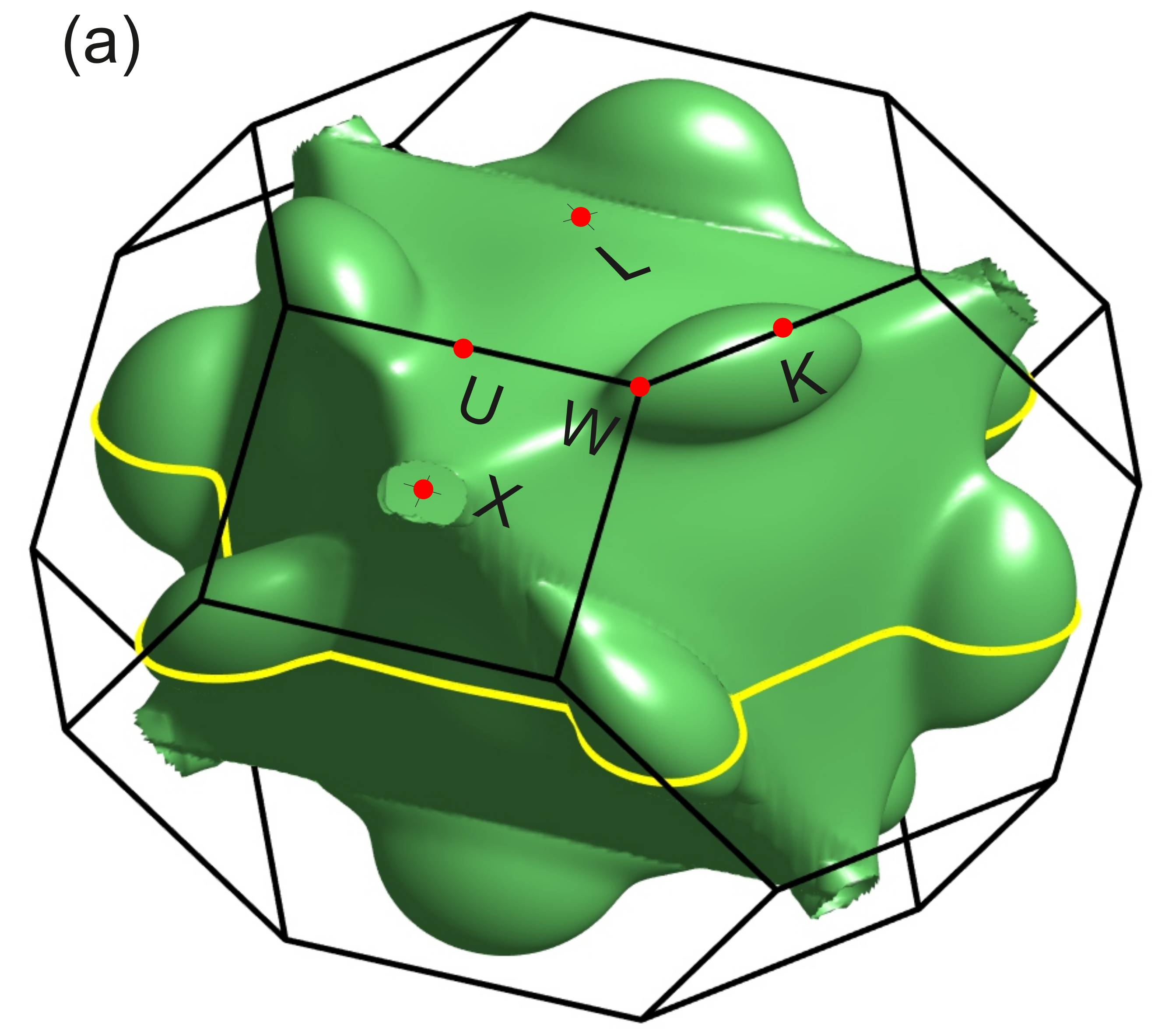} 
    \includegraphics[width=5.5cm]{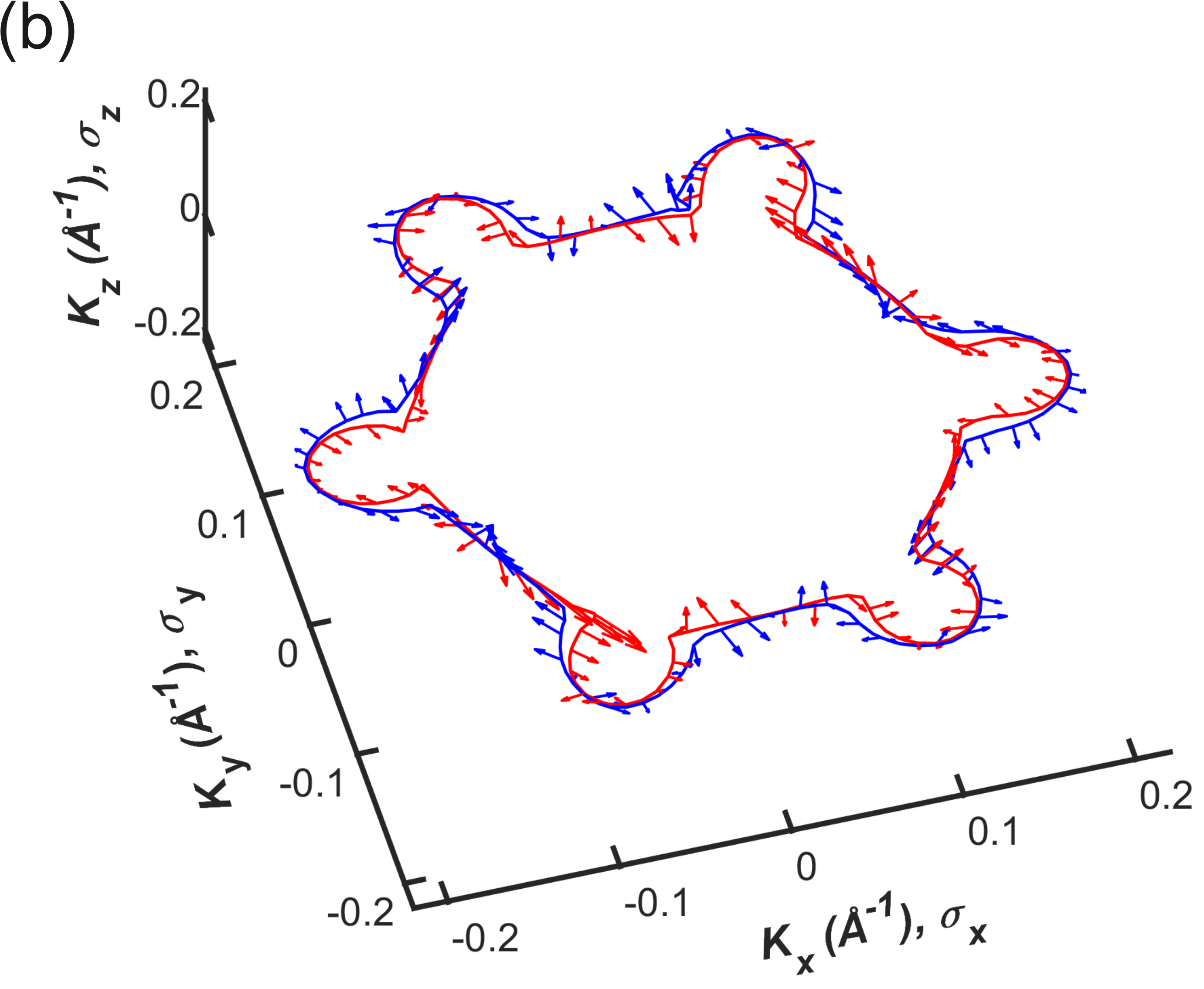}
    \caption{RuN Fermi surface: outer branch (a), cross-section (b). Averaged electron spin density along the cut lines is shown by arrows.}
    \label{fig:fermiSurface}
\end{figure}

Electron bands originating from Ru atom layers are affected by SOC, which follows from the results of RuN electronic band structure calculation (Fig.~\ref{fig:bands}). SOC leads to the splitting of the electron dispersion curves, but some electronic states are kept degenerate along $\Gamma–K$, $ \Gamma-L$ and $\Gamma-X$ lines in the Brillouin zone. 
Fermi level ($E=0$ in Fig.~\ref{fig:bands}) crosses split dispersion curves along $X-W$ line and degenerate ones along $\Gamma-K$ and $\Gamma-L$ lines, which produces two  Fermi surface inner and outer branches. The calculated Fermi surface outer branch is shown in Fig.~\ref{fig:fermiSurface} within the Brillouin zone. The yellow solid line on the surface in Fig.~\ref{fig:fermiSurface}(a) is a cross-section line corresponding $K_z=0$ cut plane. The Fermi surface cross-section is shown in Fig.~\ref{fig:fermiSurface}(b).

Electron spin density for each band and wave vector was calculated from the wave functions obtained. Spin density averaged over reciprocal unit cell is shown in Fig.~\ref{fig:fermiSurface}(b) by arrows and demonstrates the opposite spin density orientation of the both Fermi surface branches.

\begin{figure}[ht]
\includegraphics[width=8cm]{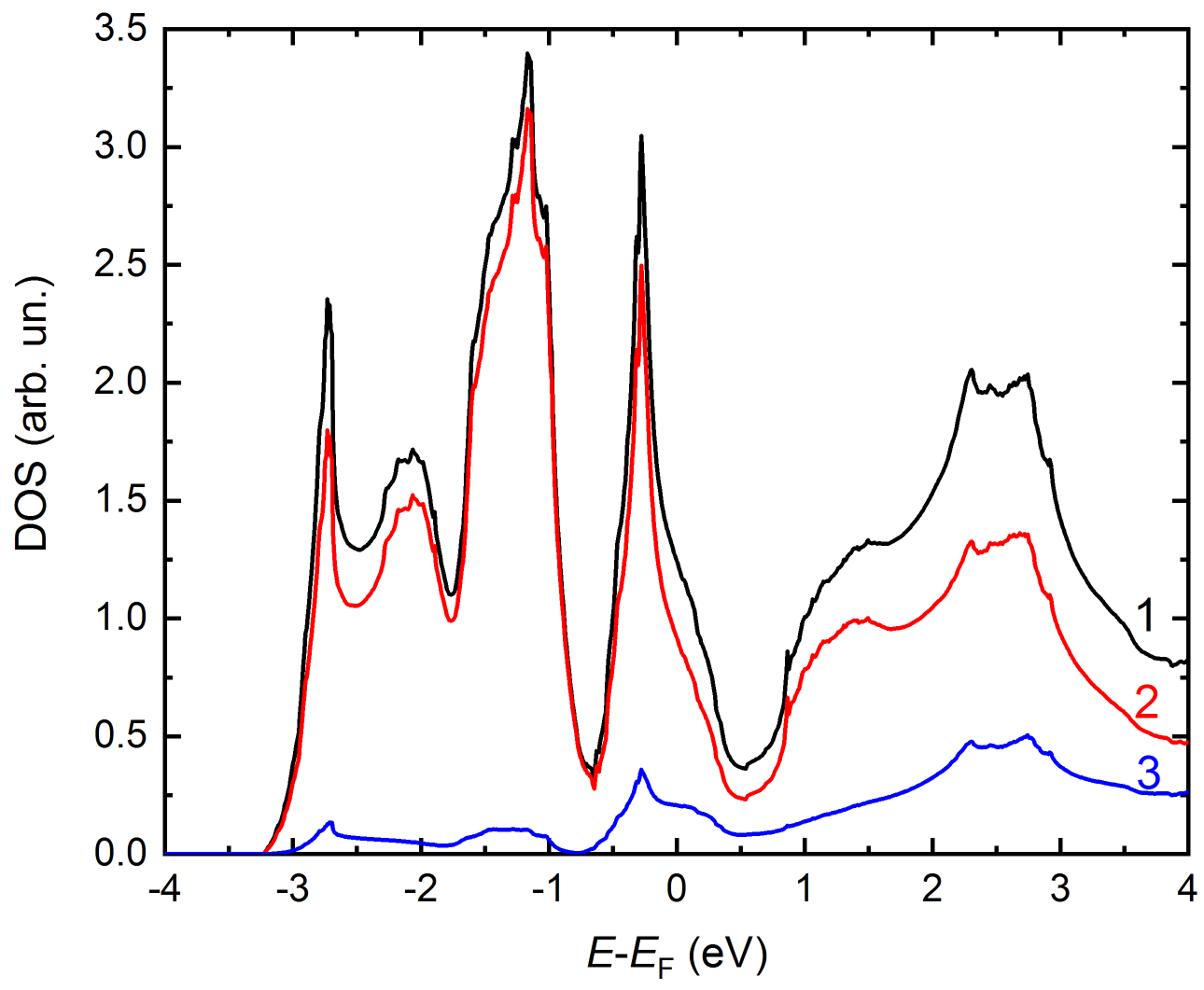}
\caption{ \label{fig:dos} RuN total (black curve, 1), Ru atom orbitale projected (red curve, 2) and N atom orbitale projected (blue curve, 3) density of states. }
\end{figure}

The density of states (DOS) presented in Fig.~\ref{fig:dos} demonstrates the metallic character of the material. It is similar to the results for the cubic F43m \cite{CHEN2010243} and tetragonal I42m \cite{Zhang2016} structures. Atom orbitale projected DOS curves show the main contribution of Ru atom orbitals (82\%) into the DOS at Fermi energy $E_F$ (red curve in Fig.~\ref{fig:dos}).

Electron-phonon coupling parameters $\lambda_{e-ph}$ and $\omega_{ln}$ \cite{Allen_1975}, which define RuN superconducting state characteristics, were calculated with the usage of Quantum Espresso package  \cite{Giannozzi_2009}. $36\times 36\times 36$ $K$-mesh was used for self-consistent electron wave functions calculation, and $6\times 6\times 6$ $Q$-mesh was used for phonon spectrum calculation. The following values were obtained: $\lambda_{e-ph}=0.55$, ${\omega}_{ln}=386$~K.

\section{Conducting and Superconducting Properties of R\MakeLowercase {u}N Films: Resistivity, Upper Critical Field and Critical Current}

\subsection{Sample parameters and measurement methods}

Sample sizes are $8\times 8$ mm$^2$ for Ru/Si film and $3.6\times 8$ mm$^2$ for RuN/SiO$_2$ and RuN/oxidised Si films. We used soldered indium contacts for resistance measurements. Contact resistance was undetectable against the  spread resistance background. The measurements were performed in the van der Pauw geometry for RuN/Si film and in the geometry with current contacts placed on the narrow ends of the substrates and potential probes attached along the wide edges of the substrates for RuN/SiO$_2$ and RuN/oxidized Si films. Measurements of the temperature dependences of resistance over a wide temperature range were carried out using a lock-in amplifier SRS 830 in a closed-cycle refrigerator RNA 10-320 (temperature range 10--300 K) and also in a homemade cryosystem based on a closed-circle refrigerator (temperature range 1.1 -- 300~K). We use helium as a heat exchange gas at a pressure of 1-10 Torr. Cooling and warming rates for were around 1.5 K/min. Critical magnetic fields $H_{c2}(T)$ and critical currents were measured in the delusion refrigerator Bluefors BF-250LD (maximum magnetic field 1~T) in the current modulation regime using Keithley 6221 current source coupled with Keithley 2182a nanovoltmeter.  When measuring $H_{c2}(T)$, magnetic field was oriented perpendicular to the films.  Critical current was measured in the point contact geometry with $\sim 100$~$\mu$m diameter indium point contact. Cooling and warming rates were around 0.005 K/min.

\subsection{Temperature Dependent Resistivity}

\begin{figure}[ht]
\includegraphics[width=8cm]{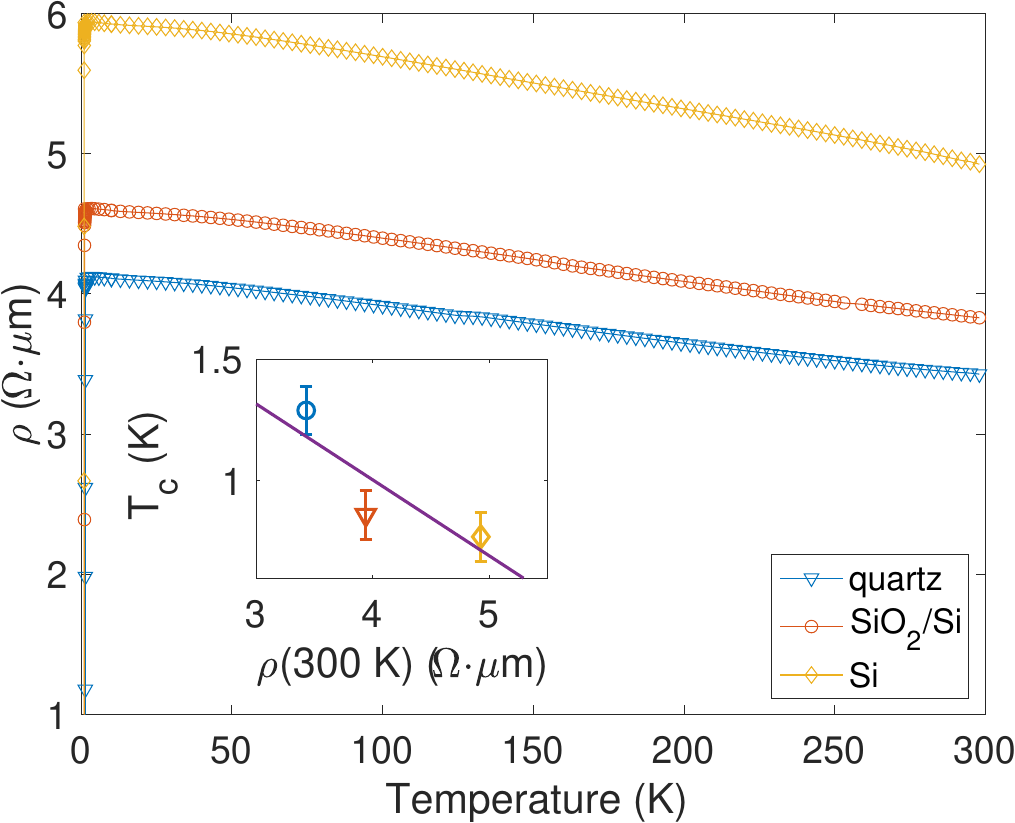}% Here is how to import EPS art
\caption{\label{fig:rho(T)} 
Temperature dependences of resistivity of RuN films grown on different substrates in a single technological process. Film thicknesses are 105~nm (see \protect{Fig.~\ref{fig:Xray23}}). Inset shows the critical temperature {\it vs.} resistivity dependence. Error bars correspond to the uncertainty associated with contact sizes. The line is a guide for the eye and corresponds to a linear root mean square approximation.}
\end{figure}

Figure~\ref{fig:rho(T)} shows resistivity data of RuN films.% grown on substrates of polycrystalline SiO$_2$ (quartz), silicon with an oxide layer obtained by thermal oxidation and silicon with natural oxide in a single technological process. 
The temperature dependencies are nonmetallic ($dR/dT<0$) for all the films, in similarity with disordered films of superconducting nitrides NbN~\cite{NbN}. The resistivity values are slightly different for the films and vary between 3.4 and 5.0~$\Omega\cdot\mu$m, which is very close to the resistivity values of disordered NbN films~\cite{NbN}.

\subsection{Critical Temperature and Upper Critical Field}

\begin{figure}[ht]
\includegraphics[width=8cm]{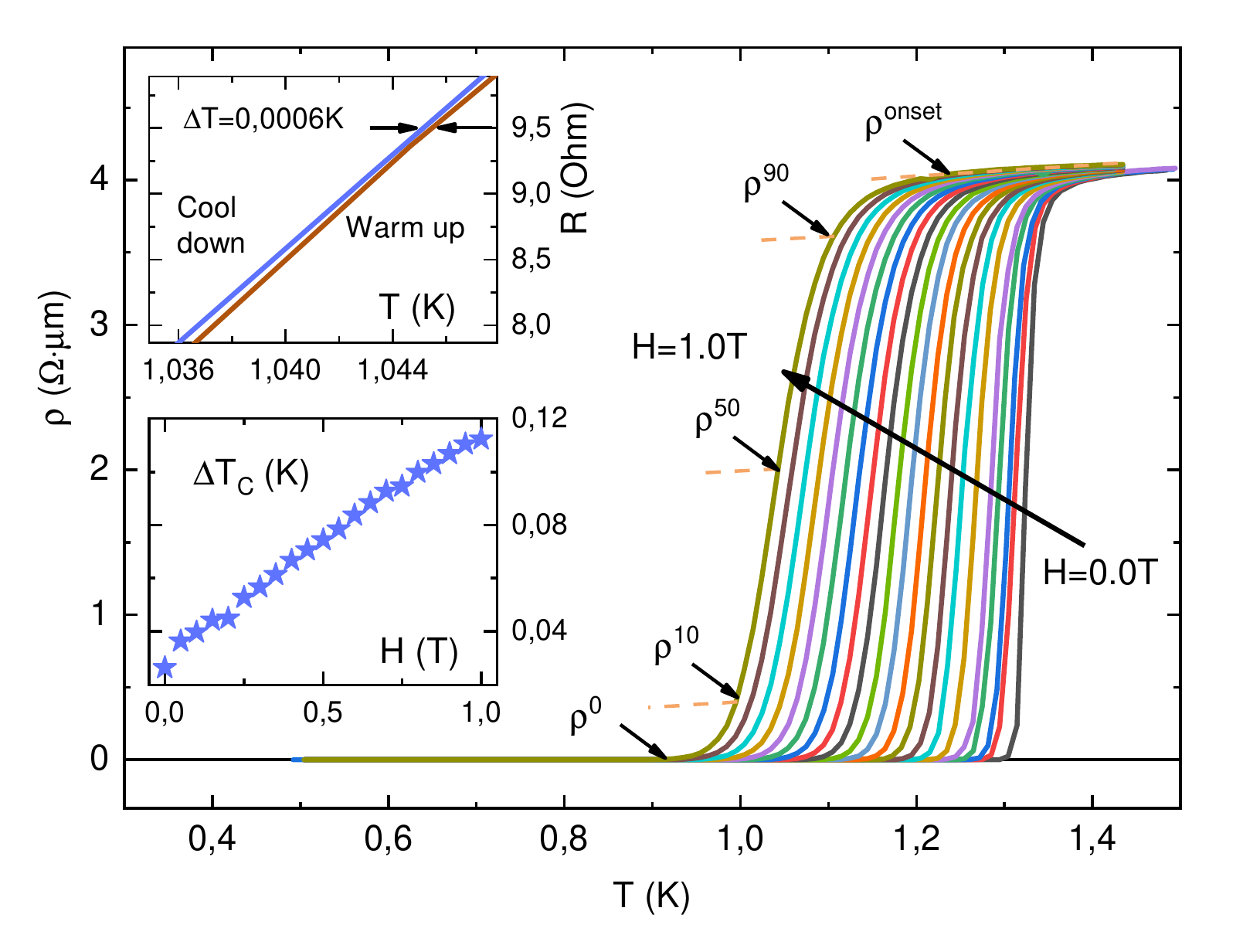}% Here is how to import EPS art
\caption{\label{fig:glass} Superconducting transitions in a RuN film on a polycrystalline SiO$_2$ substrate and their evolution with increasing magnetic field. Top inset: an enlarged section of the resistance versus temperature curve during cooling and heating. On the inset below: the width of the superconducting transition as a function of the magnetic field.}
\end{figure}

\begin{figure}[ht]
\includegraphics[width=8cm]{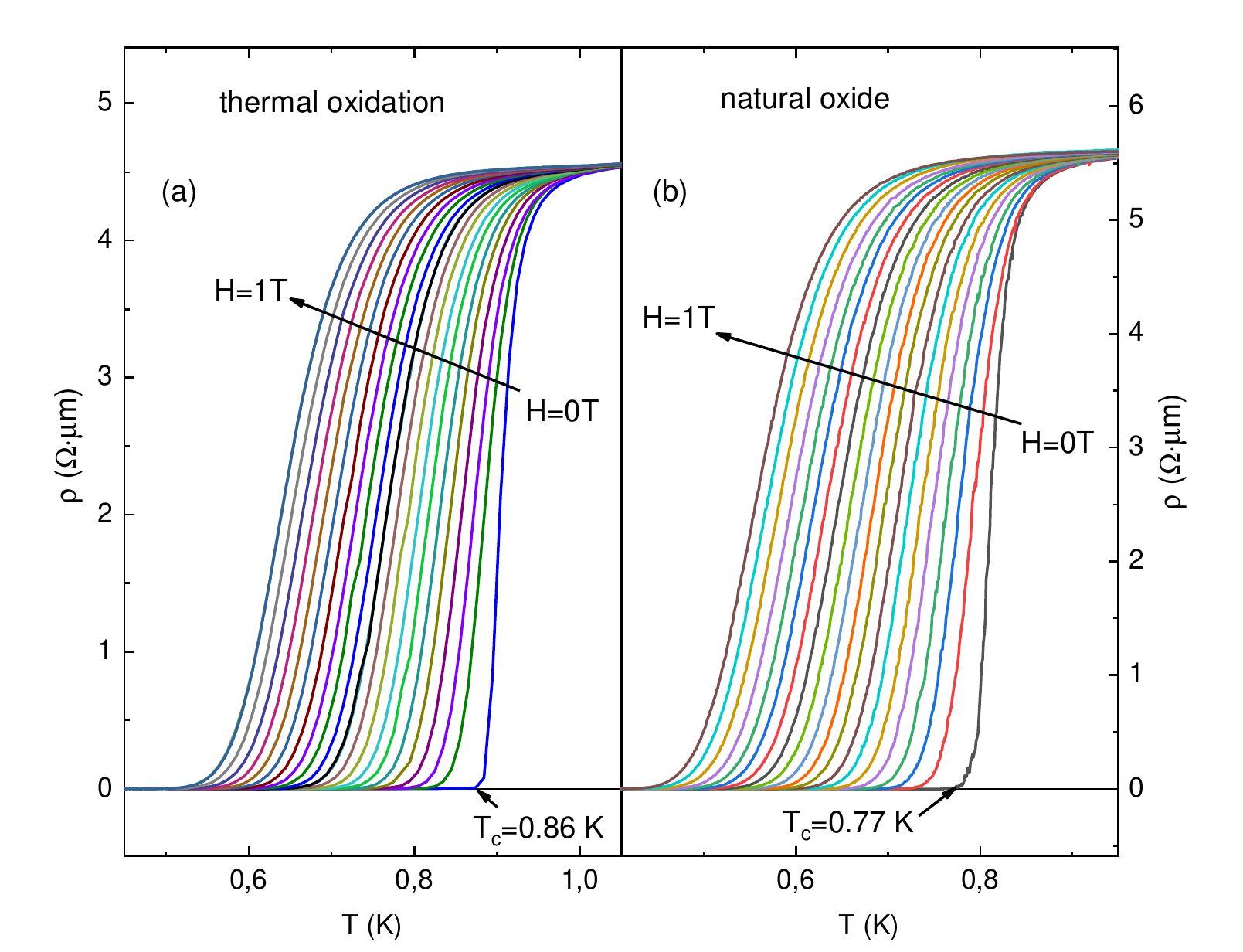}% Here is how to import EPS art
\caption{\label{fig:si_natural} Superconducting transition in RuN films on a crystalline silicon substrate with a natural oxide layer (a) and thermally oxidized surface (b) and its evolution with increasing magnetic field. }
\end{figure}

Figures~\ref{fig:glass} and \ref{fig:si_natural} show the results of measuring the resistive transition in RuN films grown on substrates of polycrystalline SiO$_2$ (Fig.~\ref{fig:glass}), silicon with an oxide layer obtained by thermal oxidation (Fig.~\ref{fig:si_natural}, left panel) and silicon with natural oxide (Fig.~\ref{fig:si_natural}, right panel) at various values of the magnetic field. In a zero magnetic field, these films exhibit a relatively narrow resistive transition at temperatures of 1.29~K, 0.86~K and 0.77~K, respectively, measured at the onset of nonzero resistance. There is anticorrelation between the specific resistance at room temperature and the critical values of the films: the higher is the specific resistance, the lower is the critical temperature (see inset in Fig.~\ref{fig:rho(T)}). The superconducting resistive transition gradually shifts to lower temperatures with an increasing magnetic field up to 1~T. It is worth noting that the width of the superconducting transition for all three samples increases significantly with increasing magnetic field, which is characteristic of hard type II superconductors, and one can expect that thermal fluctuations can play a significant role in the dissipation processes in magnetic fields.

\begin{figure}[ht]
\includegraphics[width=8cm]{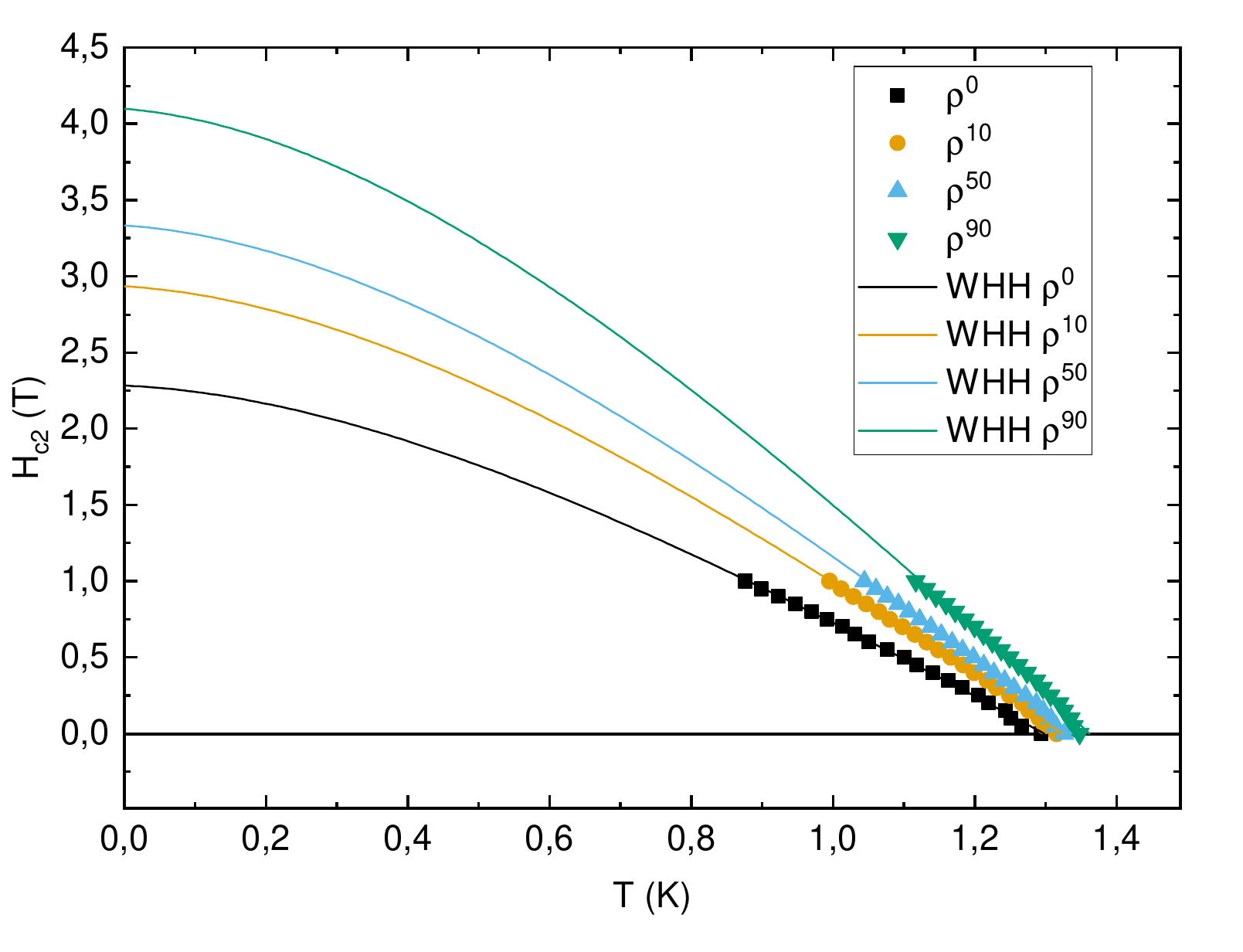}% Here is how to import EPS art
\caption{\label{fig:glassHc2} Temperature dependence of the upper critical field of a RuN film on a polycrystalline SiO$_2$ substrate for various criteria for determining its magnitude (see text). The lines show the approximation by the WHH model \cite{WHH_1,WHH_2,WHH_3} for single-zone type II superconductors.}
\end{figure}

Figure~\ref{fig:glassHc2} shows the temperature dependencies of the upper critical magnetic field $H_{c2}$ obtained from the curves shown in Fig.~\ref{fig:glass} measured at different values of the magnetic field. Four curves correspond to four standard criteria for determining $H_{c2}$ ($\rho^0$ is zero resistance, $\rho^{10}$ is 10\% of the "value" of the superconducting transition, $\rho^{50}$, $\rho^{90}$ are, respectively, 50\% and 90\% of the transition). An increase of the width of the superconducting transition with increasing magnetic field, which is characteristic of hard type II superconductors, is clearly seen. In the measured temperature range, the $H_{c2}(T)$ curves follow the WHH model  \cite{WHH_1,WHH_2,WHH_3}. From the results of approximation by the WHH model, the values $H_{c2}(0) = 2.3-4.1$~T and $\xi = 9$-12~nm for the zero resistance criterion have been obtained. 

\begin{figure}[ht]
\includegraphics[width=8cm]{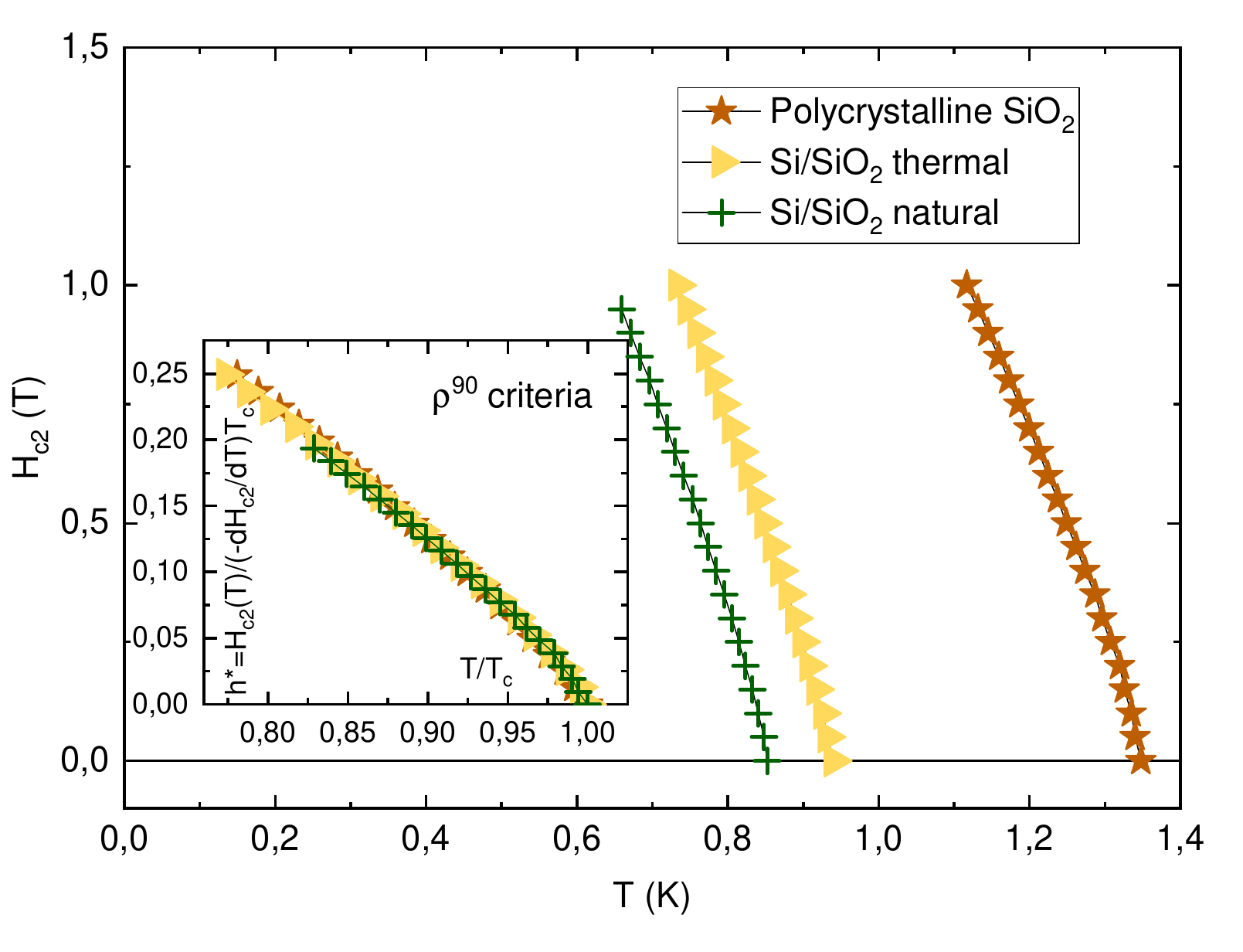}% Here is how to import EPS art
\caption{\label{fig:3Hc2} Temperature dependence of the upper critical field for three different RuN films: a polycrystalline SiO$_2$ substrate and Si/SiO$_2$ substrates with thermal and natural oxidation. Inset: temperature dependence of a reduced critical field $h^*$ for r90 criterion. }
\end{figure}

Figure~\ref{fig:3Hc2} shows the temperature dependencies of the upper critical field for all three samples $H_{c2}$ obtained from the resistive transitions in Figures \ref{fig:glass} and \ref{fig:si_natural}. For clarity, we used only one criterion - r90. The inset shows the upper critical field temperature dependence in reduced coordinates: $h^*=H_{c2}(T)/(-dH_{c2}/dT)T_c$ {\it vs.} $t=T/T_c$, so that it is clear, that even though $T_c$ and $H_{c2}$ are strikingly different for different films, the $T$-dependence is the same.

%The question arises as to why films deposited in a single technological process have such different $T_c$ values and other superconducting parameters. The results of the reflectometric study shown in Fig.~\ref{fig:Xray23} lead us to conclude that this difference results from film degradation caused by diffusion mixing of the film layer composition with the substrate. 

\subsection{Critical Current and the Energy Gap}

The ability to carry a non-dissipative current, sometimes of very high density, is one of the most valuable inherent properties of the superconductors. Not only it is an appealing attribute for practical use, but also it is a fundamental feature that gives information about the intrinsic properties of a superconductor. As was first suggested by Talantsev {\em et.al} \cite{talantsev2015universal}, the temperature dependence of self-field critical current can be used to derive the energy gap and London penetration depth of superconductors. This model was successfully used to derive fundamental superconductor parameters in several dozens of materials, starting with conventional superconductors \cite{talantsev2015universal}, copper-based HTSC \cite{talantsev2017thermodynamic}, iron-based HTSC \cite{talantsev2019p} and even most recently discovered superhydride HTSC \cite{talantsev2017london}, \cite{troyan2023non}, \cite{sadakov2023vortex}. 

For thin films (when the half thickness of a film is smaller than the penetration depth $\lambda$), the self-field critical current density, $J_c^{sf}$ in zero external magnetic field (when the magnetic field is generated only by the current itself) is related to the penetration depth as follows:\cite{talantsev2015universal},

\begin{equation} \label{eq1}
%\begin{split}
J_c^{sf} = \frac{\phi_0}{4\pi\mu_0\lambda^3}(\ln\kappa + 0.5) %\\
%\end{split}
\end{equation} 
where $\phi_0$ is flux quantum, and $\kappa$ is Ginzburg-Landau parameter, which remains almost constant under the logarithm.

Since superfluid density $\rho_s$ $\sim$ $\lambda^{-2}$, one can determine the temperature dependence of superfluid density by measuring self-field critical current density $J_c^{sf}(T)$. Because of the practical temperature independence of $\ln(\kappa$), for type-II superconductors $J_c^{sf}(T)$ is dependent only on $\lambda$, which provides a tool to extract the magnitude of the superconducting gap and even its symmetry. In particular, for $s$-wave symmetry in a single-band superconductor: 

\begin{equation} \label{eq2}
%\begin{split}
\frac{\rho_s(T)}{\rho_s(0)} = \frac{\lambda(T)^{-2}}{\lambda(0)^{-2}} = 1 - 2\sqrt{\frac{\pi\Delta(0)}{k_BT}}e^{-\Delta(0)/k_BT}%\\
%\end{split}
\end{equation}

Thus, combining equations \ref{eq1} and \ref{eq2}, we can analyze our data within a model for a type II thin film superconductor with $s$-wave single gap. 

\begin{figure}[ht]
\includegraphics[width=8cm]{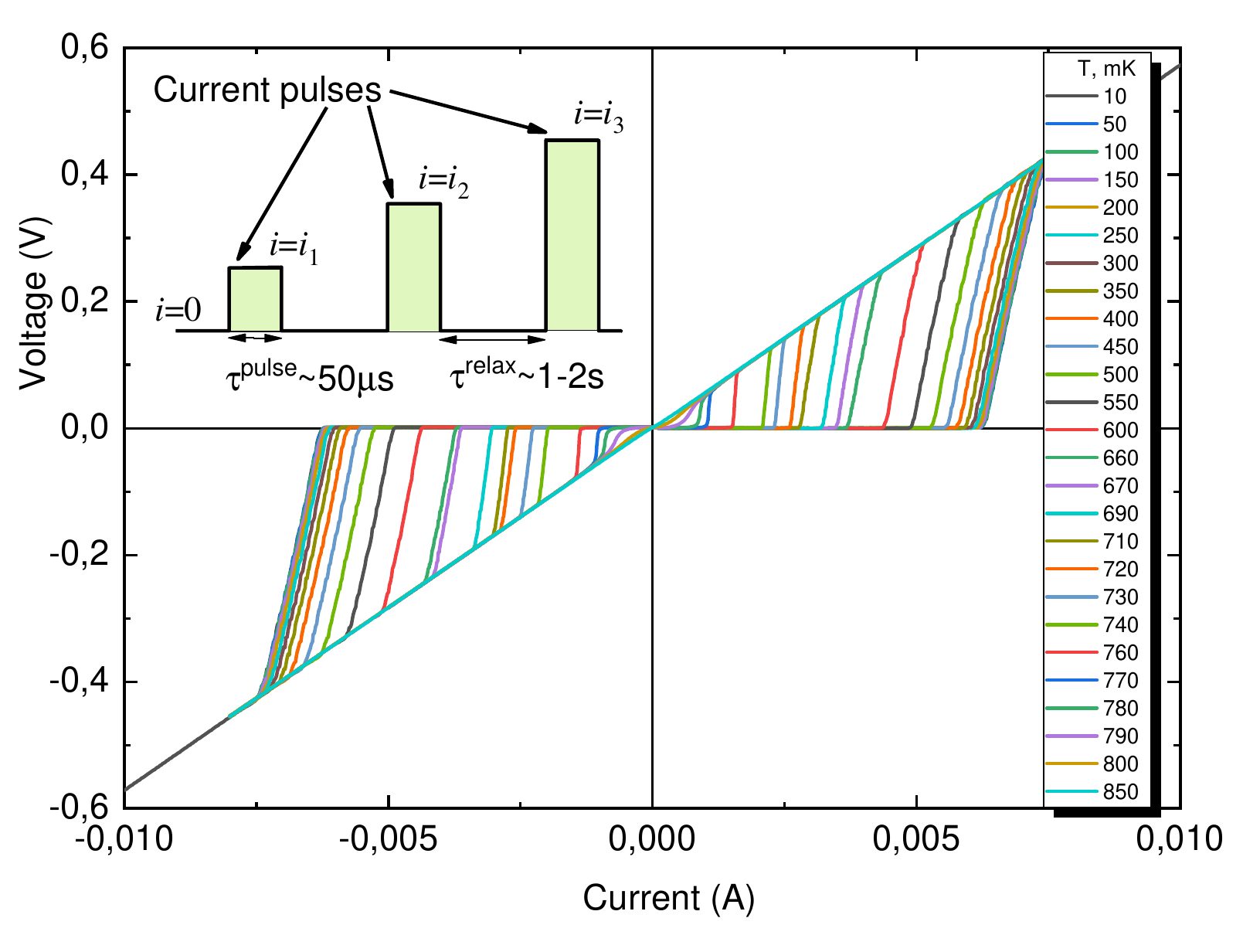}
\caption{\label{fig:IV} Current-Voltage characteristics for the RuN film on Si/SiO$_2$ substrate with natural oxidation in the temperature range of 10~mK-800~mK. The top left inset shows the principal scheme of the pulse measurements used for every I-V curve.}
\end{figure}

Figure~\ref{fig:IV} shows a set of current-voltage characteristics for RuN film on Si/SiO$_2$ substrate with natural oxidation. Measurements were carried out on a dilution refrigerator with a standard 4-probe technique. To prevent overheating the current leads, we used a high-quality pulse current source (Keithley 6221) and a nanovoltmeter (Keithley 2182a). The pulse width was as low as 50~$\mu$s. 

\begin{figure}[ht]
\includegraphics[width=8cm]{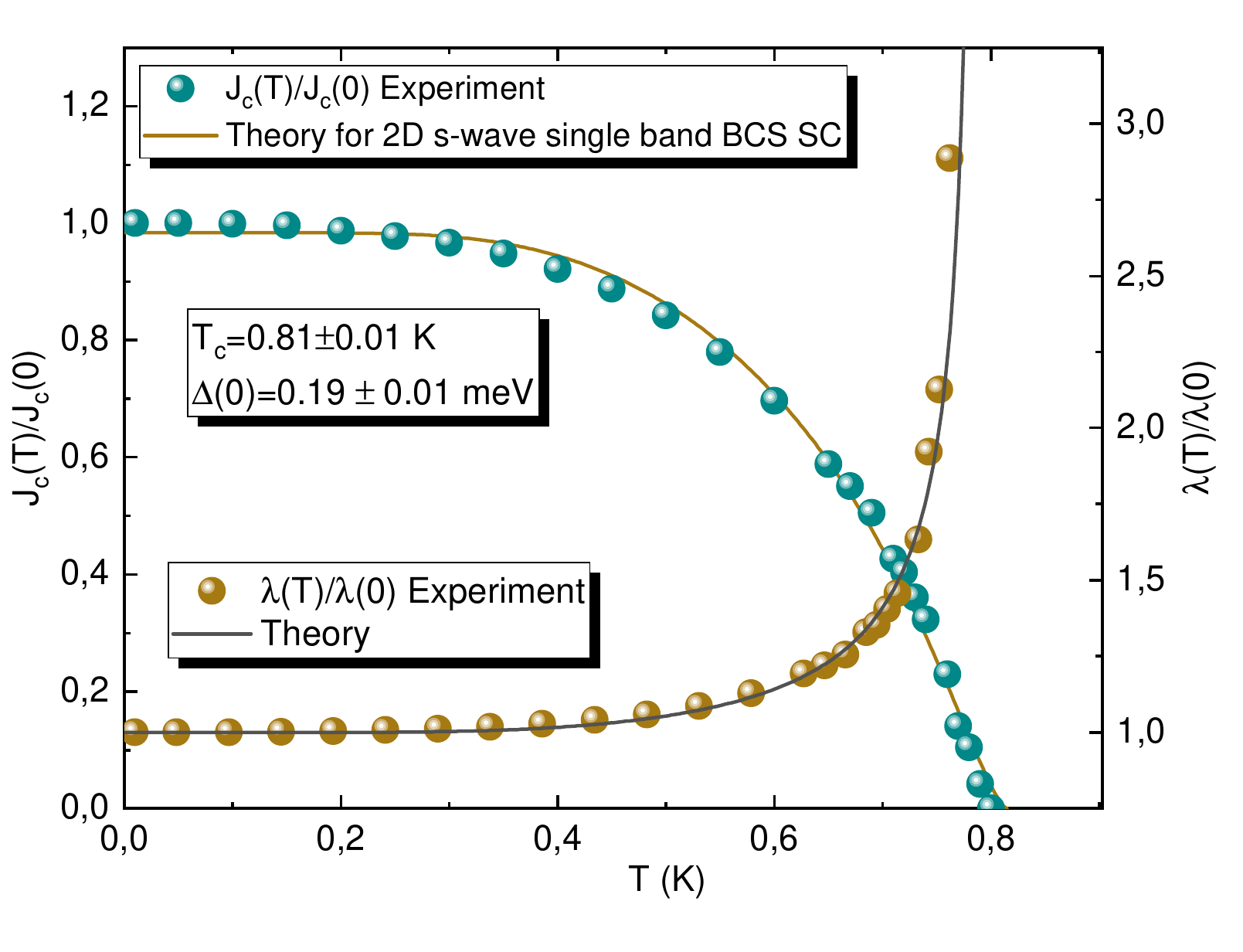}
\caption{\label{fig:Jc} Temperature dependence of the self-field critical current for RuN film on Si/SiO$_2$ substrate with natural oxidation ($J_0(0)=2\cdot 10^4$~A/cm$^2$). The dark yellow line shows a fit by the Talantsev model for a 2D s-wave single band type II superconductor.}
\end{figure}

Figure~\ref{fig:Jc} shows the temperature dependence of the self-field critical current for RuN film on Si/SiO$_2$ substrate with natural oxidation. The data points (blue symbols) were extracted from Fig.~\ref{fig:IV} by 0.5~mV criteria. We see that the Talantsev model \cite{talantsev2015universal} for 2D films with type II single gap superconductivity (light brown solid line in Fig.~\ref{fig:Jc}) fits the experimental data very well. The extracted energy gap value is $\Delta(0) =  0.19$~meV. 

\section{Discussion}

In studied RuN films, the critical temperature of the superconducting transition temperatures ranges from 0.77 to 1.29~K, depending on the substrate. The question arises as to why films deposited in a single technological process have such different  values and other superconducting parameters. The results of the reflectometric study shown in Fig. 3 lead us to conclude that this difference results from film degradation caused by diffusion mixing of the film layer composition with the substrate. In addition, there is anticorrelation between $T_c$ and the room temperature resistivity of the films (see inset in Fig.~\ref{fig:rho(T)}). This anticorrelation is very similar to that observed in another superconducting nitride NbN \cite{NbN}, where the effect is explained by differences in the degree of disorder in NbN films. In turn, since the thermal conductivity of silicon is two orders of magnitude higher than that of quartz, there is a correlation between the thermal conductivity of the substrate (and the surface temperature during film growth) and $T_c$. It is very likely, therefore, that additional substrate heating during film growth could help to improve the film quality and enhance the critical temperature.  Further study is required to verify this hypothesis.

Experimental $T_c$ lies in the interval from 1.29~K to 1.4~K for RuN films on a quartz glass substrate according to Fig.\ref{fig:glass}. On the other hand, $T_c$ can be estimated from the results of the simulation using a semi-phenomenological formula based on Eliashberg's theory \cite{Allen_1975} with calculated $\lambda$ and ${\omega}_{ln}$ parameters. The experimental $T_c$ interval can be obtained at the Coulomb pseudopotential values $\mu=0.189$-0.192.  
In these calculations, $\mu$ is considered as a phenomenological parameter. Conventional values of $\mu$ are lying in the interval from 0.1 to 0.15 \cite{Allen_1975},
so calculated $T_c$ overestimates somewhat measured one. Similar $\mu$ overestimation occurs in some transition metals as Nb and V \cite{drzazga,rietschel}. This can be explained by neglecting some additional electron-electron correlations in $T_c$ calculations. One possible mechanism of $T_c$ correction is destructive influence of the spin fluctuations \cite{rietschel}.

The transition temperature determines the upper Pauli paramagnetic limit of $H_{c2}$, which has a form  $H_p{\rm [T]}=1.86T_c$[K] for BCS-like superconductors \cite{Clogston_1962,Chandrasekhar1962}.
Measured $T_c$ values correspond to $H_p=2.4$-2.6~T. It is known \cite{Schossmann1989} that the Pauli limiting term is renormalized for strong coupling superconductors $\tilde{H}_p=(1+\lambda)H_p$. $\tilde{H}_p=3.7$-4.0~T for the samples considered here. The maximum value of $H_{c2}$ obtained from the experiment
is equal to 4.1~T, {\em i.e.} corresponds to the maximum calculated values of $\tilde{H}_p$. It should be noted that in most typical superconductors  $H_{c2}$  is less than one half of $H_p$. To explain observed  anomaly first of all we should consider a possible $T_c$ suppression owing to the either order parameter fluctuations or sample structure related effects as the effect of the film granularity and proximity effect.
The estimation of the bulk order parameter fluctuations effect is based on Ginzburg–Levanyuk criterion for dirty superconductors 
$$
		\frac{\Delta T}{T_c } = 1.6 \frac{T_c}{(k_Fl)^3 E_F},
$$
where $\Delta T$ is a fluctuation temperature interval in the vicinity of $T_c$, $k_F$ is a Fermi wave vector and $l$ is electron mean free path \cite{larkin}.
Using $E_F\approx 3$~eV and $k_Fl\sim 1$ we get $\Delta T/T_c\approx 6\cdot 10^{-3}$, so we can neglect the influence of these fluctuations on $T_c$.
Both the fluctuations in granular superconductors and proximity effect in thin film should be accompanied by $H_{c2}(T)$ curve upward curvature near $T_c$ \cite{deutscher,sidorenko}. This upward curvature is not observed in our results. Thus, we see no supporting evidence for suppression of $T_c$ by fluctuation effects or film structure effects in our measurements.

The high enough value of  $H_{c2}$ can be explained by the electron spin-orbit interaction with impurities  \cite{WHH_3}.
Fermi-level electrons are mostly related to Ru atomic orbitals according to the results of DOS computation (Fig.~\ref{fig:dos}), and an excess Ru atom concentration over N atoms (see Film composition section) provides spin-orbit scattering centers.   Moreover, the structure 
of RuN has no inversion symmetry, so we can't exclude a mixing of singlet and triplet pairing in the presence of spin-orbit interaction, which increases the upper limit of $H_{c2}$ as well \cite{Smidman2017}. Further study is required to reveal the nature of high value of $H_{c2}$.

\section{Conclusion}
In conclusion, we observed superconductivity in RuN films obtained by reactive magnetron sputtering on Si, oxidized Si, and silica glass substrates. The critical superconducting transition temperature ranges from 0.77 to 1.29 K depending on the substrate and is anticorrelated with the resistance of the films at room temperature. The crystal lattice is cubic with the following parameters: $a=b=c=4.559$~\AA~for RuN/SiO$_2$,  $a=b=c=4.536$~\AA~for RuN/Si, and $\alpha =\beta =\gamma =87.96^\circ$ for both films. DFT calculations, together with the results of temperature-dependent self-field critical current measurements, prove that RuN is a single-gap superconductor with  $\Delta(0) =  0.19$~meV. The upper critical magnetic field $H_{c2}(T)$ depends on the substrate and, within the framework of the WHH model, corresponds to $H_{c2}(0)=2.3$–4.1~T and $\xi = 9$–12~nm for the coherence length. $H_{c2}(0)$  is near the paramagnetic limit of the strong coupling case.

\section{Data availability}
Data will be made available on a reasonable request.

\section{Declaration of Competing Interest}
The authors declare that they have no known competing financial interests or personal relationships that could have appeared to influence the work reported in this paper.

\begin{acknowledgments}
The work was supported by the Russian Science Foundation (project \# 21-72-20114). The studies were carried out using the equipment of the Central Collective Use Center of the Lebedev Physical Institute of RAS. The computations in this study were performed using computational resources at the Joint Supercomputer Center, Russian Academy of Science.
\end{acknowledgments}
    
\bibliography{RuN_refs}% Produces the bibliography via BibTeX.

\end{document}